\documentclass{aastex}

\slugcomment{DRAFT - To appear in ApJ}

\shortauthors{Ardila, Basri}
\shorttitle{Balmer range}

\begin{document}

\newcommand \vsini{$v$sin$i$~}
\newcommand \caII{\ion{Ca}{2}~}
\newcommand \heI{\ion{He}{1}~}
\newcommand \liI{\ion{Li}{1}~}
\newcommand \hal{H$\alpha$~}
\newcommand \hbeta{H$\beta$~}
\newcommand \hgam{H$\gamma$~}
\newcommand \rsun{R_\odot}
\newcommand \msun{M_\odot}
\newcommand \lsun{L_\odot}
\newcommand \kms{km s$^{-1}$~}
\newcommand \degs{$^\circ$}

\title{The Balmer Wavelength Range of BP Tauri}

\author{David R. Ardila\altaffilmark{1}, Gibor Basri\altaffilmark{1}}
\altaffiltext{1}{Astronomy Dept., Univ. of California, Berkeley, CA 94720,\\e-mail: ardila@garavito.berkeley.edu, basri@soleil.berkeley.edu}

\begin{abstract}
We have analyzed all the observations of BP Tauri taken by the International Ultraviolet Explorer in the low resolution ($\bigtriangleup \lambda \sim 6 \rm{\AA}$), long wavelength (from $\lambda=1850 \rm{\AA}$ to $\lambda=3350 \rm{\AA}$) range. This dataset contains 61 spectra. We observe variability in the ultraviolet continuum of $\bigtriangleup m_{cont.} \sim 1 $ magnitude and variability in the \ion{Mg}{2} line flux of $\bigtriangleup m_{\rm{Mg II}} \sim 0.8 $ magnitudes. Moreover, these spectra do not show any correlation between the continuum flux and the \ion{Mg}{2} line flux, thus resolving a standing controversy in the literature concerning the origin of the \ion{Mg}{2} line flux. There is no correlation between the color temperature of the UV continuum and the average value of its flux. Using models of the accretion process developed by Calvet \& Gullbring (1998), we obtain energy fluxes, accretion spot sizes, and accretion rates from the IUE observations of BP Tauri. We find average energy fluxes of $5.0\ 10^{11} \rm{ergs\ cm^{-2}\ s^{-1}}$, average spot sizes of $4.4\ 10^{-3}$ times the stellar surface, and average accretion rates of $1.6\ 10^{-8} {\rm{\msun/yr}}$. Our analysis shows that the particle energy flux and the UV flux in the stellar surface are proportional to each other. Most strikingly, we observe a correlation between accretion rate and spot size, with the spot size increasing as the square of the accretion rate. Based on the results of a simulation, we conclude that geometrical effects (i.e. the varying inclination of the spot with respect to the observer) are not enough to account for this effect. Current models of the accretion process fail to reproduce such an effect, suggesting the need of using more realistic descriptions of the stellar field when treating magnetospheric accretion. There may also be an unmodelled efficiency factor that determines how matter is loaded into the field lines. Non-dipole fields, geometry, oblique shocks and the possibility of ``limb brightening'' should be taken into account when creating models and explaining the results of observations of T-Tauri stars. 

\end{abstract}

\keywords{stars: pre-main-sequence --- stars: individual(BP Tauri) --- space vehicles --- ultraviolet: stars}

\section{Introduction}

The International Ultraviolet Explorer (IUE) was operational for almost 20 years and took over 107000 images of a wide array of objects. In particular, it observed about 130 T-Tauri stars (TTSs) in the low resolution ($\bigtriangleup \lambda \sim 6 \rm{\AA}$), long wavelength range (from $\lambda=1850 \rm{\AA}$ to $\lambda=3350 \rm{\AA}$). This dataset constitutes the most complete set of observations of TTSs in the Balmer range.

Now that the IUE Final Archive has been completed and all the data is in the public domain\setcounter{footnote}{0}\footnote{http://archive.stsci.edu/iue/} we have undertaken the project of exploring all this dataset. In this paper we will examine the observations of BP Tauri (also known as HBC 32 and HD 281934). A complete discussion of the full dataset will appear elsewhere.

We have chosen this star in particular because of the large number of observations by IUE: there are 61 low resolution spectra. They will allow us to study correlations between different physical aspects of the T-Tauri phenomenon over a long period of time.

BP Tauri is a single (Ghez, Neugebauer \& Matthews 1993), Classical T-Tauri star (CTTS) located in L1495 (Lynds 1962) in the Taurus-Auriga complex of dark clouds at a distance of 140 pc (Elias 1978). The star has a constant radial velocity close to that of its molecular surroundings and a projected equatorial velocity of $v {\rm sin}i < 10 $ km/s (Hartmann et al. 1986; Basri \& Batalha 1990). The spectral type of BP Tauri corresponds to a K5-K7 dwarf with a strong absorption in the lithium resonance line (Duncan 1991; Hartigan et al. 1989). The mass is $M=0.5 - 0.8 \msun$ (Gulbring 1994; Gullbring et al. 1998).

Photometric observations (Vrba et al. 1986; Bouvier, Bertout \& Bouchet 1988; Simon, Vrba \& Herbst 1990 - SVH90 - and the observations summarized by Rydgren et al. 1984) have shown that BP Tauri has a large infrared emission and is variable in all optical bands, with a range in B of up to 3 magnitudes (Herbig \& Bell 1988). Rapid fluctuations on time-scales of minutes have been observed by Schneeberger, Worden \& Africano (1979).

These fluctuations make it difficult to extract a possible photometric period. No period at all (Bouvier et al. 1988, Simon et al. 1990) and periods of 6.1 days (Simon et al. 1990), 7.6 days (Vrba et al. 1986), 7.7 days (Osterloh, Thommes \& Kania 1996), 8.3 days (Richter et al. 1992), and $\sim 10$ days (G\'omez de Castro \& Franqueira 1997 - GdCF97) have been proposed. The fact that BP Tauri grows redder as it becomes fainter has been used to explain this variability in the period  in terms of a hot spot on the surface of the star (see for example SVH90). In a small fraction of CTTSs, spots cooler than the photosphere have also been observed, but no evidence of cool spots exists for BP Tauri (Gullbring et al. 1996). Variations in the measured period of BP Tauri might correspond to the hot spot (or spots) moving to different latitudes in a differentially rotating star. An inclination of $i\sim 30^o-50^o$ for BP Tauri has been derived on the basis of the measured $v \sin i$ and the longest measured periods (SVH90; Gullbring 1994). The variability in the photometric period can also be explained by considering a beat frequency between the Keplerian frequency of clumps of material circling the star and the rotation rate of the stellar surface (Bouvier et al. 1999; Smith, Bonnell \& Lewis 1995; Smith, Lewis \& Bonnell 1995).

Measurements of spot sizes in this and other TTSs give values that vary between a fraction of a percent to almost 40\% the visible stellar surface (Vrba et al. 1986; Vrba et al. 1993; Herbst, Herbst \& Grossman 1994; Gullbring 1994; Bouvier et al. 1995;  Fern\'andez \& Eiroa 1996).

As has been observed in many CTTSs, the spectrum of BP Tauri appears veiled by variable continuum excess emission. This is the origin of the photometric variability. The optical veiling has been observed to be between 30 and 75\% the value of the photospheric flux (Basri \& Batalha 1990; Valenti, Basri \& Johns 1993; Hartigan, Edwards \& Ghandour 1995; Gullbring et al. 1998). 

Some spectral lines (notably, the Balmer series and the \ion{Mg}{2} line) show evidence of blueshifted and redshifted absorption (see for example, Johns \& Basri 1995ab; GdCF97).  It has also been noted that line emission and the continuum level are sometimes correlated (GdCF97; SVH90). We will explore one of these correlations in Section 3.

These observations can be framed in terms of the current paradigm for CTTS.  As we understand it today, a CTTS consists of a late-type pre-main-sequence star with a magnetic field, surrounded by an accreting, dusty disk. At the distance from the star at which the torque due to the magnetic field equals the viscous torque in the disk, the disk is truncated. Some material from the disk will then be captured by the magnetic field (see for example K\"onigl 1991) and some will escape as a wind (Shu et al. 1994). Collisional coupling between ions and neutrals is such that even the neutral gas component follows the magnetic field lines (Martin, 1996). At the stellar surface, the material will have supersonic velocity. The simplest models assume that this velocity should be the free-fall velocity ($v_{ff}\sim 300 \ \rm{km \ s^{-1}}$, Calvet \& Gullbring 1998) but more complicated models slow down the material to a fraction ot this velocity (Ostriker \& Shu 1995). Nevertheless, a shock will occur some distance above the star. The shocked gas will heat up the surface of the star and a hot spot will be seen. Emission from this hot spot is thought to be responsible for the visible and ultraviolet excess observed in CTTS spectra (Valenti et al. 1993; Calvet \& Gullbring 1998; Lamzin 1998). Furthermore, emission from the whole accretion flow is believed to be responsible for other features of the TTS spectra, in particular, the strong emission in the Balmer lines (Hartmann, Hewett \& Calvet 1994; Johns-Krull \& Basri 1997) and some atomic lines (Batalha et al. 1996; Beristain, Edwards \& Kwan 1998). In this model redshifted absorption features would correspond to material falling into the accretion spot. Blueshifted absorption features would correspond to absorption by an optically thin wind. It is unclear at this point what fraction of the excess emission (if any) is due to the wind.

In the context of the magnetospheric accretion model, the ultraviolet is a particularly interesting range. Emission from the underlying star is low compared to that of the veiling continuum. This means that, even though is not possible to measure the veiling directly (as it is possible in the visible range), conclusions derived from the ultraviolet analysis are not very sensitive to errors in the spectral type.

In this paper we will explore the IUE dataset for BP Tauri. Relatively little work has been done on the individual IUE observations and their relation to the accretion paradigm. Here we assume the excess ultraviolet continuum emission to be caused by accretion. We start with a discussion of the raw data. A more complete discussion (including many more stars) will appear soon, in a paper dealing with the full low-resolution IUE dataset. In Section 3, we will use our dataset to study the correlation between the UV continuum and the flux in the \ion{Mg}{2} blend. To understand the IUE data to the fullest we need a physical model. Therefore in Section 4 we will use the Calvet \& Gullbring (1998) (CG98) description of the accretion shock to interpret our spectra in terms of accretion rates. Section 5 and 6 contain the analysis and the conclusions.

\section{Observations}

We searched the IUE Merged Log for observations of TTS taken with the long wavelength cameras (LWP and LWR, from $\lambda=1850 \rm{\AA}$ to $\lambda=3350 \rm{\AA}$) and obtained spectra for a total of 131 different TTS. For BP Tauri we found 61 low resolution ($\bigtriangleup \lambda \sim 6 \rm{\AA}$) spectra. A summary of the observations is shown in Table 1. The oldest spectra comes from April 1979 and the most recent from January 1992. There are clusters of observations taken in October 1986 (SVH90) and January 1992 (GdCF97).

All the spectra used in this paper have been processed by the New Spectra Image Processing System (NEWSIPS). For an explanation of this system and the differences with IUESIPS, see Nichols \& Linksky (1996). For each observation the spectrum has been sky-subtracted by using the background in the camera. NEWSIPS spectra come with an estimate of the error at each wavelength. This error is derived empirically for each camera by measuring the scatter in the flux around the mean in the background regions of several hundred images (Garhart et al. 1997). As such, it is an estimate of the instrumental, statistical and background noise for each spectrum. NEWSIPS also provides a ``quality'' flag at each wavelength of each spectrum. This flag indicates certain special events, such as a saturated pixel, cosmic ray or the presence of a calibration mark.

Figure 1a. shows the mean and the normalized variance (Johns \& Basri 1995b) of all spectra. The feature at $\lambda\simeq3050\rm{\AA}$ is a reseaux mark. The central line is the blending of the \ion{Mg}{2} h \& k lines.  Figure 1a shows that the edges of the spectra vary the most, whereas the center is very constant. The noise of the spectra also increases towards the edges (due to the reduced sensitivity of the Vidicon detectors used by IUE) but not enough to account for the value of the variance.

Figure 1b shows two extreme manifestations of the spectra (not corrected for reddening). The lower one (lwp18976, solid line) shows strong emission of C II] and Fe II multiplets. For a more exhaustive description of the spectral features present, see Imhoff \& Appenzeller (1987)  and Brown, de M. Ferraz \& Jordan (1984). In the upper one (lwp09256, dash-dot line), those emissions are very difficult to distinguish. These spectral changes are typical of CTTS and are probably due to inhomogeneities in the accretion flow and/or changes in the orientation of the star as it rotates. The temperature of the upper spectrum is almost a thousand Kelvin higher than that of the lower spectrum (Table 1).

We have de-reddened all the spectra using $A_V=0.51$ magnitudes (Gullbring et al. 1998). This reddening determination is based on the shape of the optical spectrum and not only on broad band photometry. It is therefore more reliable than previous values of $A_V$.

In addition to its science objects, IUE also observed a suite of standard stars. We have compared spectral characteristics of BP Tauri with two of these standards (Table 1). The color temperature of the BP Tauri spectra was evaluated using the continuum fitting procedures described in Sections 4.2 and 5.3. HBC 399 (also known as V 827 Tau) was the only WTTS observed by IUE with enough signal-to-noise to measure its color temperature. It is classified as K7, M0 and has an \hal equivalent width of $-1.8\AA$ (Herbig \& Bell 1988). We take the reddening to be $A_V=0.28$ magnitudes (Kenyon \& Hartmann 1995).  IUE also observed main-sequence stars including 61 Cyg B, with spectral type K7V. According to the SIMBAD database its parallax is $\sim0.28$ arc-sec, which means that the reddening is negligible. As Table 1 shows, the continuum level and the flux in the Mg II line in BP Tauri are one and three orders of magnitude larger than those in HBC 399 and 61 Cyg B respectively. This observation confirms the fact that the intrinsic photospheric continuum of the star is invisible in the Balmer range. As shown in Section 4 this ultraviolet excess can be understood as being due to the accretion process.

\section{The continuum and the \ion{Mg}{2} blend}

As mentioned in the introduction, relationships between line emission and continuum level have been published by a number of authors. The study of these relationships is interesting because they illuminate what fraction of the observed line emission is due to the accretion process, as opposed to the wind or the stellar photosphere. For BP Tauri in the ultraviolet, two main articles reaching opposite conclusions have been published. These articles use subsets of the entire IUE dataset for BP Tauri.

From six spectra taken in 1986, SVH90 reached the conclusion that variations in the flux of \ion{Mg}{2} blend can be correlated with continuum variations. They also looked at \ion{O}{1} ($1300 \rm{\AA}$), \ion{Si}{2} ($1810 \rm{\AA}$) and \ion{C}{2} ($1335 \rm{\AA}$) and reached the same conclusion. This would seem to indicate that the bulk of the emission in those lines comes from a process directly related to the formation of the accretion spot. From a set of twenty-two spectra taken in 1992, GdCF97 reached the conclusion that the  \ion{Mg}{2} blend, \ion{Si}{2} and \ion{C}{4} ($1550 \rm{\AA}$) light curves are dominated by short timescale fluctuations, whereas \ion{O}{1} and \ion{He}{2} ($1640 \rm{\AA}$) are well correlated with variations in the UV continuum. \ion{Mg}{2}, \ion{Si}{2} and \ion{C}{4} are collisionaly excited lines, but \ion{O}{1} and \ion{He}{2} can be excited by recombination process. 

The IUE data utilized in this article includes all the data used by SVH90 and GdCF97 and much more and therefore it can resolve the issue of whether or not the \ion{Mg}{2} line flux correlates with the UV continuum. Following GdCF97, we have measured the UV continuum variations in a window of $50 \rm{\AA}$ width centered at $2900 \rm{\AA}$. This region of the spectra stands between two groups of Fe II lines (Brown et al, 1984) and therefore it samples closely the real continuum of the star. 

Disregarding those spectra for which the continuum or the \ion{Mg}{2} blend are affected by ``extraordinary'' events (defined as having a ``quality'' flag different from zero), we have 43 measurements. The dispersion in the continuum flux measured in this way, is $\bigtriangleup m_{cont.} \sim 1 $ magnitudes. The dispersion in the flux of the line is $\bigtriangleup m_{\rm{Mg II}} \sim 0.8$ magnitudes. Figure 2 shows that, indeed, no correlation exists between the flux in the blend and the UV continuum flux. 

The difficulty of interpreting low-resolution observations of the \ion{Mg}{2} line is compounded by the fact that high resolution observations, as described by GdCF97 show that the line has a strong wind component. If part of the line is created in the accretion flow and part in an extended wind, a measurement of the low resolution flux is sampling to different regions of the matter flow (see Section 5.3).

We have also compared the color temperature of the continuum (with the continuum as found in Section 4.2) and the amount of continuum flux (Figure 3). The errors in the temperature are $\sim10\%$. As Figure 3 shows, no correlation exists between color temperature and flux in the continuum. This implies that there is no correlation between the flux in the line and the color temperature. We defer further discussion of this Figure until Section 5.3.

\section{Modeling the ultraviolet continuum}

In order to interpret the spectra, a physical model is needed; we have decided to use the work of CG98. A full description of the model can be found there. We describe only those aspects of it directly relevant to our data. In this section we will find the accretion rate, the energy flux and the spot size indicated by each of our spectra. 

\subsection{The Model}

CG98 construct a model in which the observed spectral excess in CTTS is the sum of optically thick emission coming from the heated photosphere, and optically thin emission coming from the pre- and postshock regions. The shock is assumed to be parallel to the surface of the star. The heated postshock material radiates mainly in X-rays. Half of that radiation is reprocessed into optical and UV wavelengths in the photosphere; the other half is absorbed in the preshock region and reprocessed in optical and UV wavelengths. A fraction of this emission will be re-emitted towards the star and will contribute to heating the photosphere, whereas the remainder will travel away from the star.

The amount of observed radiation at each wavelength depends on two factors. The first factor is the strength of the potential depth of the star, M/R, where M is the mass and R is the radius of the star. This ratio determines the temperature of the postshock gas and therefore, everything else being equal, the hardness of the incident radiation.  We have assumed $\rm{M/R}=0.25 \msun/\rsun$ for BP Tauri (Gullbring et al. 1998). The second factor is the kinetic energy flux carried by the accretion flow. From mass conservation we have that the density of the accretion flow is $\rho={\dot M}/(A_{int-spot}*v_{ff})$, where ${\dot M}$ is the accretion rate, $v_{ff}$ is the free-fall velocity of the material. $A_{int-spot}$ is the intrinsic area of the spot, equal to the observed area divided by $\cos\theta$, the inclination of the spot with respect to the line of sight. To consider inclination effects explicitly, we write  $\rho={\dot M}\cos\theta/(A_{spot}*v_{ff})$, where now $A_{spot}$ is the observed spot area. We then write the kinetic energy flux (or the particle flux) as:
\begin{equation}
{\mathcal{F}}={\frac{1}{2}}{\rho} {{v_{ff}}^3}=9.8\  10^{10} ({{\dot M}\cos\theta \over {10^{-8} \msun \ yr^{-1}}})({M \over {0.5 \msun}})({R \over {2\rsun}})^{-3}({f \over {0.01}})^{-1} \\ \rm{\ ergs \ cm^{-2} \ s^{-1}}
\end{equation}

in which $v_{ff}\simeq 300\ \rm{km\ sec^{-1}}$, the free-fall velocity from a truncation radius equal to $\rm{R_T}=5 R_{*}$ (Meyer, Calvet \& Hillenbrand 1997), $f$ is the apparent fractional size of the accretion region or spot, $A_{spot}=f 4 \pi R_{*}^2 $ and $\cos\theta$ is the inclination of the spot with respect to the line of sight. The factor of $\cos\theta$ is obtained by assuming that the star has only one spot. If the region producing the emission is not one spot but a set of spots or a deformed ring (Mahdavi \& Kenyon 1998) $\cos\theta$ should be interpreted as the cosine of the mean inclination of the region responsible for the emission. In the following analysis we do not find the intrinsic accretion rate but ${\dot M}\cos\theta$, the ``projected'' accretion rate. Similarly, the value that is found for $f$, the spot size, corresponds to the observed spot size and not the intrisic one. Both are related by $f=f_{int}\cos\theta$.

CG98 assumes that the truncation radius $\rm{R_T}$ remains constant. Modeling by a number of authors (e.g. K\"onigl 1991, Ghosh 1995, Ostriker \& Shu 1995) suggests that $\rm{R_T}\ \alpha \ {\dot M}^{-2/7}$. This result can be obtained in a number of ways, but the simplest is by realizing that it is the only dimensionally correct dependence between $\rm{R_T}$ and $\rm{\dot M}$. Furthermore, in most models the truncation radius affects the size of the spot in such a way that a large $\rm{R_T}$ implies a small spot (see Section 5.6). If we were to assume a varying truncation radius, the factor $\zeta=1-R_*/R_T$ would appear multiplying the accretion rate. As the truncation radius depends only weakly on the accretion rate, we will assume $\zeta$ to be constant .

CG98 divides the emission accretion zone into three regions: Shock/postshock, preshock, and heated atmosphere. In the shock and postshock region a number of simplifying assumptions are made to solve the mass, momentum, and energy conservation equations: the flow is assumed to be plane-parallel and strictly along field lines (i.e., one-dimensional) perpendicular to the stellar surface. Energy transport by conduction is neglected, as are instabilities on the flow and effects of the magnetic field on the dynamics of the gas. For the boundary conditions the shock is assumed to be strong. The temperature structure towards the stellar surface is calculated until a minimum is reached. The emission from the accretion column is calculated with CLOUDY.

The preshock region is assumed to be illuminated by the radiation field provided by the postshock calculation. The small size of this region compared to the stellar radius is used to justify an assumption of constant density  through it.

The stellar atmosphere is assumed to be semi-infinite, plane-parallel, and in radiative equilibrium. The emission from it is the sum of the unperturbed stellar photosphere, plus reprocessed incident radiation. For $\rm{T_{*}}=4000$ K, $\rm{M}=0.5 \msun $, $\rm{R}=2 \rsun$ and $\log {\mathcal{F}}= 11.5$, the effective temperature of the continuum at $\tau_{Ross}=2/3$ is then $\rm{T_{eff}}\sim 6000 K$. The output of the model is a grid of the excess spectra each of which is characterized by a different value of $\log {\mathcal{F}}$.

Through CG98's model, the excess Balmer continuum is shown to be due to optically thick emission from the heated atmosphere and optically thin emission from the pre- and postshock regions. The former dominates at high energy fluxes ($\log {\mathcal{F}} > 11 $) with the total spectrum becoming more like blackbody emission. The optically thin emission becomes more important as the energy flux decreases. The Paschen continuum is always dominated by optically thick photospheric emission. We refer the reader to the original paper for further details.

\subsection{Fitting the continuum}

To use CG98 models, the first step is to find the UV continuum. We have only used IUE spectra in which there are regions with S/N$>$2.5. As this region differs for each image, the useful wavelength coverage of each spectrum varies slightly, but in general, the useful range is reduced from the nominal IUE coverage to $\lambda\simeq2400 \rm{\AA}$ to $\lambda\simeq 3100 \rm{\AA}$. The signal-to-noise constraint reduces the number of useful spectra to 45. We have developed an automated procedure that eliminates from each spectrum the regions with the resonant \ion{Mg}{2} blend and the \ion{Fe}{2}\ uv 1 blend (2771.3 to 2821.3 \rm{\AA}\ and 2576 to 2636 \rm{\AA}, respectively). For the remaining data, the bottom of the flux in a $\bigtriangleup \lambda \sim10\rm{\AA}$ wide box is found. This defines the continuum. The error in the continuum determination is found from the variations in the flux within this box. The continuum points found with this procedure are binned until we have 20 data points.

As Figure 1b shows, there are certain regions of the spectrum that sample the apparent continuum. Our procedure makes sure that the continuum spline fit goes through certain points like $2300 \rm{\AA}$, $2650 \rm{\AA}$, $2900 \rm{\AA}$ and $3000 \rm{\AA}$.

\subsection{The Temperature of the Underlying Star}

To obtain the spectrum of the excess emission we must first characterize the spectrum of the underlying K7 star. As shown in Table 1 the color temperature of the standard stars of the same spectral type as BP Tauri are lower than the canonical 4000 K of a K7 star. This is because the UV photosphere is optically thicker than the visible one and therefore it is situated close to the temperature minimum of the star. If we assume that BP Tauri is intrinsically like HBC 399, the color temperature of its photosphere should be $\sim3200$K. It has been shown that the ultraviolet color temperature of early G stars is $\sim80\%$ of their effective temperature  (Haisch \& Basri 1985). It is also shown by CG98 that the ratio of Balmer to Paschen continua for a $\rm{{T_{eff}} =4000 K}$ star is better reproduced by assuming that the underlying star has a temperature of 3580 K in the Paschen continuum. They attribute this to the possibility that accretion is occurring into a cool spot. What the temperature of this spot is in the Balmer range cannot be estimated without further modeling. 

Therefore, for the purposes of calculating the excess emission, we assume that the underlying star has a temperature of $3580$K. This estimate is consistent with the color temperature of the standards and allows us to use CG98 models. We will examine the consequences of this assumption in Section 5.4.

\subsection{Finding the Energy Flux and the Spot Size}

Once the underlying star has been subtracted, the resulting observed excess emission is compared with model spectra given by CG98. To perform this comparison, we obtain the mean value and the slope (assuming a straight line) of the logarithm of the observed spectrum. These two parameters have zero covariance and so any correlation between them is not due to the fitting procedure. 

In CG98 models, the slope of the predicted excess emission for a fixed value of $\rm{M/R}$  depends only on $\log \mathcal{F}$. By fitting the continuum we are able to associate a value of  $\log \mathcal{F}$ with each observed spectrum. When our values for $\log \mathcal{F}$ extend beyond their published grid, we have linearly extrapolated their models. Once the value of  $\log \mathcal{F}$ is found, we find the relative size of the spot $f$ by comparing the mean value of the theoretical excess spectrum to that of the observed excess spectrum. 

In this way we obtain the values for $\log \mathcal{F}$ and $f$ quoted in table 2. Errors in $f$ are $\sigma_{f}/f\ \sim10$\% and are a combination of the errors in the value of the flux at each wavelength, and errors in the determination of the continuum.  Errors in $\log \mathcal{F}$ are $\sigma_{\log \mathcal{F}}/{\log \mathcal{F}}\ \sim1$\%. With Equation 1 we obtain ${\dot M}\cos\theta$, the accretion rate uncorrected for projection effects. With this model it is impossible to obtain the ``unprojected'' accretion rate ${\dot M}$, unless $\cos\theta$ is known: $\mathcal{F}$ is independent of $\theta$ but the observed spot area $f$ is proportional to $\cos\theta$ and so their product is proportional to ${\dot M}\cos\theta$. The errors in ${\dot M}\cos\theta$ are $\sigma_{{\dot M}\cos\theta}/{\dot M}\cos\theta \ \sim30\%$. All errors include an error in $\rm{A_V}$ of 0.05 magnitudes.

\section{Analysis}

The average size of $f$ is $\overline{f}=4.4\ 10^{-3}$ with a dispersion of $\sigma= 5\ 10^{-3}$. The average of $\log \mathcal{F}$ is $\overline{\log \mathcal{F}}=11.7$, and the dispersion is $\sigma=0.2$ (see Figure 4a and 4b). It should be noted that the value of $f$ is a lower limit to the real spot size, given that what one observes is the projected spot area. Also, because of the period changes mentioned in the introduction, we may not be observing the same spot (or set of spots) all the time, but their size and distribution over time. Figure 4c shows the distribution of ${\dot M}\cos\theta$. The average is $\overline{{\dot M}\cos\theta}=1.6\ 10^{-8} \msun \ \rm{yr^{-1}}$, and the dispersion is $\sigma \sim 0.8 \ 10^{-8}\msun/yr$. The dispersion of $f$ is larger than the observational errors, but the dispersions in $\log \mathcal{F}$ and ${{\dot M}\cos\theta}$ are consistent with the errors.

As mentioned in the introduction, a wide range of spot sizes have been identified by other authors. We find spot sizes ranging from 0.06 to 3\% the size of the visible star disk. The average accretion rate we observe agrees with the most recent determinations for BP Tauri, based on single optical spectra: $\dot M \simeq 2-3 \ 10^{-8} \msun \ \rm{yr^{-1}}$ (Gullbring et al. 1998, CG98).

The logarithm of the energy flux, $\log \mathcal{F}$, does not depend on the inclination angle $\theta$. Therefore, the variations observed in Figure 4b are due to variations in ${\dot M}$ and/or the intrinsic size of the accretion spot. The dispersion in $\log \mathcal{F}$ implies a fractional variation of $\sim 50 \%$ in ${\dot M}/f_{int}$, where $f_{int}$ is the intrinsic (i.e. deprojected, $f_{int}=f/\cos\theta$) spot size. This dispersion is consistent with the errors in $\mathcal{F}$. Variations in ${\dot M}$ cannot be decoupled from those in $f_{int}$. What can be said is that changes in the accretion rate ${\dot M}$ or $f_{int}$ are less than 50\% during the time spanned by the observations. This upper limit puts constraints on the inhomogeneities of the disk (that may contribute to the variations in the accretion rate) and in the geometry of the stellar field.

\subsection{Variable Extinction}

Recently, Bouvier et al. (1999) reported on photometric observations of the CTTS AA Tauri that can be interpreted as periodic occultations of the stellar photosphere by a warp in the accretion disk. One of the effects of this warp is to act as a source of variable reddening. This mechanism might also play a role in the IUE observations of BP Tauri. To find out if that is the case, we performed the following analysis. The procedure of finding which model corresponds with an observed spectrum has three free parameters: reddening, $f$ and $\log \mathcal{F}$. In this paper we assume a fixed reddening and find the values of $f$ and $\log \mathcal{F}$ that match the observed spectrum to the models. We can also assume a fixed spot size and find the values of $\log \mathcal{F}$ and reddening that match the observed spectra. We did so, assuming $f=4\ 10^{-3}$. For the cases in which this procedure succeeded in finding a $\log \mathcal{F}$ and $A_V$ that fit the data, we obtained a large range of values for $A_V$, from 0.4 to 10 magnitudes, but most of them bigger than 1. The largest value that we found in the literature for a reddening determination of BP Tauri was 0.85 (Valenti et al. 1993). Therefore, most of the reddenings determined in this way do not fit the optical data. Photometric observations in various bands simultaneously have also failed to detect color changes in BP Tauri analogous to those reported by Bouvier et al. (1999) for AA Tauri. We therefore conclude that the changes in $\log \mathcal{F}$ and $f$ are not due to changes in reddening.

\subsection{The accretion luminosity versus the continuum flux}
 
Figure 5 shows a tight correlation between ${\dot M}\cos\theta$ and the continuum flux, which implies that the observed (projected) accretion rate determines the brightness of the UV continuum. We can convert the observed UV flux to luminosity and the model accretion rate to accretion luminosity. We obtain
\begin{equation}
{\log (L_{acc}/ \lsun)}={(1.1\pm0.1)*\log (L_{2900 \AA}/ \lsun)-(0.5\pm0.05)} 
\end{equation}

A similar relationship has been found by GC98 and Gullbring et al. 1997, who observe a very clear correlation between the accretion luminosity and the brightness in the U-band, based on individual observations of various CTTSs in the optical. The empirical relation they find is
\begin{equation}
{\log (L_{acc}/ \lsun)}={1.09*\log (L_U/ \lsun)+0.98} 
\end{equation}

in which $L_{acc}$ is the accretion luminosity (which is proportional to the accretion rate) and $L_U$ is the unreddened luminosity of the star in the U-band once the intrinsic (photospheric) luminosity has been subtracted. 

An independent way to verify the accretion rates obtained here is to use Equation 3 as to predict accretion luminosity and then compare it with the accretion luminosities from Table 4. Given the rapid photometric variations of CTTSs, one needs to use concurrent $L_{acc}$ and $L_U$ to test this relationship. We searched the photometric database maintained by William Herbst\footnote{http://www.astro.wesleyan.edu/~bill/research/ttauri.html\#ov2} described in Herbst et al. (1994) and found two measurements of the U-magnitude close enough to the IUE measurements: RJD=5615.51 (IUE at 5615.85) and RJD=5620.99 (IUE at 5620.97), which imply $\log (L_U/\lsun)=-1.88$ and $\log (L_U/\lsun)=-2.05$ respectively. These determinations assume $L_{BP Tauri}=0.925\lsun$ (Gullbring et al. 1998). Using Equation 3, the predicted values are $\log (L_{acc}/\lsun)=-1.072$ and $\log (L_{acc}/\lsun)=-1.25$ respectively. We observe $\log (L_{acc}/\lsun)=-1.101$ and $\log (L_{acc}/\lsun)=-1.303$, by interpolating Figure 5. This is consistent with Equation 3 within the errors of our data. Admittedly, two points out of 45 is a very limited sample, but the fact that they obey Equation 3 lends support to our determinations of the accretion rate.

\subsection{The Line Flux and the Continuum Temperature}

Figure 3 shows that no correlation exists between the continuum temperature and the flux in the continuum. Within the context of CG98 model it is possible to understand this result. The Balmer continuum is the sum of optically thick emission from the heated photosphere and optically thin emission from the pre-shock region. Only when the value of ${\mathcal F}$ is very high does the observed color temperature approach the temperature of the heated photosphere. The flux in the continuum at $2900 \rm{\AA}$ depends, again in the context of CG98's model, not only on the spot size, but also on the value of ${\mathcal F}$. Two spectra with the same accretion spot size but different $\mathcal F$ will have different values of continuum flux. Therefore, the color temperature is a measure of two different things, the spot size and the energy flux.

We do not observe any correlation of the \ion{Mg}{2} line flux with $\log {\mathcal{F}}$, the accretion rate ${\dot M}\cos\theta$, or the spot size $f$. 

It has been argued by Hartmann (1990) that strong emission lines (like \hal and \ion{Mg}{2}) in the spectra of CTTS are formed in an extended envelope. His argument is based on the similar widths of the strong lines. Table 1 shows that the \ion{Mg}{2} line is very bright in BP Tauri, compared with that of non-CTTS of the same spectral type. There have also been unsuccessful attempts to model the \ion{Mg}{2} line blend as purely chromospheric (e.g. Calvet, Basri \& Kuhi 1984). All these arguments point to the conclusion that the \ion{Mg}{2} line blend is a consequence of the accretion process. The lack of correlation with the accretion rate maybe the result of time delay: if the \ion{Mg}{2} line is created in the funnel flow one would expect a delay of $\sim5$ days between a spectral manifestation in \ion{Mg}{2} and the landing on material on the star. An analogous delay is expected for the wind component of the line. 

On the other hand GdCF97 suggest that magnetospheric flaring may give origin to increased \ion{Mg}{2} emission. Such flaring is present in the model by Shu et al. (1994). If this is the case, it is unclear whether or not any time delay would correlate the line flux with the accretion rate. The flaring of the magnetosphere maybe be causally unrelated to the process of accretion. If this is the case we may be seeing a Mg II line corresponding to more than one source: a wind component, a funnel component and a magnetospheric component.

As the model used here to find the accretion rate is sensitive only to regions below the accretion shock, the lack of correlation of the \ion{Mg}{2} line with the accretion rate is telling us that the dominant component of the line is not created in the region below the accretion shock.

\subsection{The Accretion rate and the spot size}

Figure 6a shows that there is a correlation between the spot size, $f$, and the accretion rate, ${\dot M}\cos\theta$. The slope of the points in Figure 6a is such that ${\dot M}\cos\theta\  \alpha\ f^{0.5\pm0.05}$.

The correlation between ${\dot M}\cos\theta$ and $f$ implies (via Equation 1) a correlation between $\log {\mathcal{F}}$ and $f$, as shown in Figure 6b. Here ${\mathcal{F}}\ \alpha\ f^{-0.5}$. Assuming a constant accretion rate of ${\dot M}\cos\theta=1.6\ 10^{-8} \msun \ \rm{yr^{-1}}$, we obtain the solid line shown in Figure 6b, which shows that smaller spots have under-average ${\dot M}\cos\theta$ and larger spots have over-average ${\dot M}\cos\theta$. This merely reflects the fact that the observed (projected) accretion rate is not constant.

As Equation 2 shows, the observed (in the detector) ultraviolet flux, in $\rm{ergs/sec \ cm^2}$ is such that $F_{UV}\ \alpha \ {\dot M}\cos\theta$. This implies, from Figure 6a, that $F_{UV}\ \alpha \ f^{0.5}$. The flux on the surface of the star is proportional to the detected flux divided by the emitting area. Therefore, at the stellar surface, $F^{sur}_{UV}\ \alpha \ f^{-0.5}$ and the particle flux and the ultraviolet flux are proportional to each other. 

The proportionality between particle and surface UV flux is implicit in CG98 models because a large particle flux produces a large UV flux. What needs an explanation is the physical origin of only one of the correlations ($F^{sur}_{UV}\ \alpha \ f^{-0.5}$, ${\mathcal{F}}\ \alpha\ f^{-0.5}$ or ${\dot M}\cos\theta\  \alpha\ f^{0.5}$). Once one is explained, the others follow.

CG98 observe values of energy flux and spot sizes suggesting that, for increasing mass accretion rate, material will fall onto the star over a larger surface area, while the energy flux ${\mathcal{F}}$ remains more or less constant. They reach this conclusion by single optical observations of 15 CTTSs (Figure 6c). We also find that a bigger mass accretion rate implies a bigger spot size, but the energy flux that we calculate does not remain constant. While there is a tendency in CG98's data towards larger accretion spots with larger area (as implied by the constant ${\mathcal{F}}$), this tendency is not strong. Most of their stars lie on the locus defined by the BP Tauri observations and therefore, their data could be construed as confirming the analysis done here. However, their stars are not a homogeneous population, as they encompass a range of radii and masses.

The correlations from Figure 6 are somewhat sensitive to the estimate of the temperature of the underlying star. A higher temperature star  will produce smaller inferred accretion spots with higher $\log {\mathcal{F}}$. For the 3580 K, the temperature that we have chosen here, the ratio between the observed UV continuum flux and the photospheric flux is $\sim30$. For 4000 K this ratio $\sim10$. For the 4000K case, the accretion spots are 0.3 times the size of the spots obtained with the 3580K case. The values of $\log {\mathcal{F}}$ and $\dot{M}\cos\theta$ are 1.04 times and 0.6 times the values of the 3580K case. The shape of the correlations from Figure 6, including slopes and overall values, does not change, within the errors in the calculated quantities. In other words, even though a given spot will change size when the assumed temperature of the underlying star changes, the overall relationship between spot size, energy flux, and accretion rate does not change.

\subsection{Simulating the Correlation between Accretion Rate and Spot Size}

A possible explanation for the correlation from Figure 6a lies in the geometry of the accretion region. As the spot size, $f$, and the accretion rate, ${\dot M}\cos\theta$, are both proportional to the geometric factor, $\cos\theta$, one would expect a correlation between them. When the projected size of the accretion spot is large (large $f$ due to a small inclination angle) we should see large projected accretion rates. The larger range of the spot sizes ($\sim1.6$   dex) compared to the accretion rates ($\sim1.2$ dex) could be explained by different intrinsic distributions.  

To decide whether or not the correlation of Figure 6a is due to geometrical effects we have perfomed a simulation. We assume that the situation is one in which we are looking at an opaque disk from an inclination angle $i$. The spot has the colatitude angle $\theta_{col}$. As all the spectra in our data set show ultraviolet excess, one always see the spot. This gives the constraint that $i\le \pi/2-\theta$. Let $\hat{\mathbf{i}}$ be the unit vector pointing from the center of the star to the observer. Then $\hat{\mathbf{i}}=(\sin(i),0,\cos(i))$. Let $\hat{\mathbf{n}}$ be the unit vector perpendicular to the spot area. Then $\hat{\mathbf{n}}=(\sin(\theta_{col}) \cos(\omega t),\sin(\theta_{col}) \sin(\omega t),\cos(\theta_{col})$, where $\omega=2\pi / 8$ rad per day, the angular frequency of the star, assumed to have a period of 8 days. Therefore $t$ maps the phase in which we are observing the star. The cosine of the angle between the spot and the observer (which we have called here $\theta$) is given by $\cos\theta=\hat{\mathbf{i}} \cdot \hat{\mathbf{n}}=\sin(i)\sin(\theta_{col})\cos(\omega t)+\cos(i)\cos(\theta_{col})$. We then generate uniform random points between 0 and 8 for $t$. This procedure should produce the observed distribution of $\cos\theta$.

We assume that ${\dot M}$ and $f_{int}=f/\cos\theta$ have normal distributions.  The simulation therefore has 6 parameters: the means and sigmas of the normal distributions, the colatitude angle of the spot and the inclination of the system. For a given set of parameters, we randomly choose 2000 points for ${\dot M}$, $f_{int}$ and $\cos\theta$ and multiply ${\dot M}$ by $\cos\theta$ and $f_{int}$ by the same $\cos\theta$. The values of $f_{int}\cos\theta$ and ${\dot M}\cos\theta$ are constained to be larger than the minima of their distributions. In other words, we are assuming that the accretion rate and the intrinsic size are uncorrelated and that any correlation comes from the $\cos\theta$ that multiplies both quantities.

We do not find any combination of parameters that will satisfy all the constraints of the problem: Figures 4a, 4b and 6a. All the simulations that produce a correlation between  ${\dot M}\cos\theta$ and $f$ give a slope close to 1 for the log-log plot of the two quantities. The width of the gaussians determine the scatter of this correlation. This means that the basic hypothesis of the model (that ${\dot M}$ and $f_{int}=f/\cos\theta$ are uncorrelated gaussians) is not verified by the simulation.

Dendy, Helander \& Tagger (1988) have considered the possibility that the intrinsic distribution of accretion rates should be a power law. They arrive to this conclusion based on models of self-organized criticality. The range of accretion rates found here is not large enough to test this hypothesis. 

While it is clear that a full understanding of Figure 6a should include the effects of changing inclinations of the spot as the star rotates, the fundamental explanation is not geometry. We have to look for other mechanisms.

It is possible that some of the inferred changes in energy flux are due to ``limb brightening''. Assuming that spot size does not change for short cycles, a small spot size would imply a high spot inclination, and therefore we would observe higher regions of the spot (farther away from the center of the star), which the models predict have higher temperatures (Lamzin 1998). Therefore, these higher regions will look like regions of high energy flux. One would therefore expect a correlation between $\log {\mathcal{F}}$ and $f$, as observed here. If ``limb brightening'' were indeed an important effect, the slope of the spectra would not measure the value of $\mathcal F$ but the temperature of that layer, which would change for different orientations. The CG98 models assume that the spot is observed from directly from above, and therefore it is difficult to account for this effect. Resolving this issue will have to wait for more detailed models.

\subsection{Theoretical Expectations of the Relationship between Spot Sizes and Accretion Rates}

The total size of the accretion region is determined by geometry of the stellar magnetic field. An accretion spot of variable size would therefore imply that the field itself is changing. So far, the most sophisticated idea of the accretion process is the so-called 'X-wind' model (Shu et al. 1994). This model has been developed for a magnetic dipole aligned with the disk, even though most authors agree that such geometry is an oversimplification of the more complicated real geometry of the field. In the X-wind model, the fractional coverage of the accretion spot with respect to the area of the star (the quantity we have called here $f$) is given by $f=\sqrt{1-R_*/R_X}-\sqrt{1-1.5*R_*/R_X}$ (Ostriker \& Shu 1995), where $R_X$ is the radius of the 'X-point', that coincides to a few percent with the corotation radius of the star and with the truncation radius $R_T$. Furthermore, $R_X\ \alpha \ {\dot M}^{-2/7}$. For $R_*/R_X\sim0.2$ as assumed in this paper, $f\sim6\%$. Let's assume the accretion rate increases by a factor of 2, from the value that produces by $R_*/R_X\sim0.2$. $R_X$ decrease bys a factor of 0.8 and then  $R_*/R_X\sim0.24$. Then  $f\sim7\%$. According to Figure 6a, $f$ should increase by a factor of 4 when the accretion rate increases by a factor of 2. The increase in spot size with accretion rate is in the right direction but is not fast enough to explain the correlation of Figure 6a.

The model by Mahdavi \& Kenyon (1998)  uses a tilted dipole and consider the accretion as occurring onto a thin ring on the surface of the star. For $R_*/R_T\sim0.2 \ - \ 0.3$  and a dipole inclined by 60 degrees with respect to the plane of the accretion disk, this model predicts $f\sim2 \ - \ 4\%$, and $f\ \alpha \ {\dot M}^{2/7}$ which is again a very slow increase in the right direction.

It is also clear that the problem remains if we include explicitly the varying truncation radius in the correlation from Figure 6a: $f\ \alpha \ ({\dot M}\cos \theta)^2 (1-R_*/R_T)^2$. The increase in the accretion rate dominates over the slow decrease in the factor $\zeta=(1-R_*/R_T)$.

The root of the problem is probably in the assumption of a magnetic dipole as the magnetic field of the star. Even though this is a good assumption at large distances from the star it is clear that it is an oversimplification if one wishes to explain the geometry of the accretion spots, which depends on the field close to the star. 

In Ostriker \& Shu (1995) and Mahdavi \& Kenyon (1998), the predicted accretion spots are on the upper tail of our distribution. This has also been observed by CG98, which conclude that the accretion process has to be inhomogeneous. If not all the available area is covered at the same time, the observed accretion spot will be a fraction of the available area. 

Independently of these considerations one could argue that, according to Figure 6a, matter loading on the field lines is not very efficient. We see that the accretion rate increases only with the linear dimensions (e.g. the radius or the circumference) of the accretion spot: ${\dot M}\sim f^{0.5}$. So a bigger area produces a more diluted accretion, and therefore, a weaker surface flux $F^{sur}_{UV}$. Another way of saying this is that the spot size is very sensitive to changes in accretion rate.
 
The models that we use here are not complete: they assume a  shock parallel to the surface of the star, where the real shock may be oblique. They also assume free-fall velocity, which, as Ostriker \& Shu (1995) show, is not the most appropiate hypothesis. The inclusion of these two effects would increase the observed spot size, but it is unlikely that they will change the slope of Figure 6a. To rule out this possibility, more complete simulations are needed.

In summary, the IUE data on BP Tauri puts a strong constraint in any model that seeks to explain the size of accretion spots. Accurate modeling should consider the geometry of the observations (i.e. inclination angle, spot latitude), a more realistic geometry for the magnetic field, oblique shocks, the real velocity of the material falling onto the star, and possibly the effects of limb brightening. There may also be an unmodelled efficiency factor that determines the way in which flux tubes are actually filled with material, in such a way that the accretion rate scales as the radius of the tube.

\section{Conclusions}

We have analyzed all the observations of BP Tauri taken by the International Ultraviolet Explorer in the low resolution ($\bigtriangleup \lambda \sim 6 \rm{\AA}$), long wavelength range (from $\lambda=1850 \rm{\AA}$ to $\lambda=3350 \rm{\AA}$). We found 61 useful spectra. 

Compared with spectra of standard stars, all the BP Tauri spectra show strong ultraviolet excess and color temperatures around $8000$ K. Using 43 spectra, we observe variability in the ultraviolet continuum of $\bigtriangleup m_{cont.} \sim 1 $ magnitude and variability in the \ion{Mg}{2} line flux of $\bigtriangleup m_{\rm{Mg II}} \sim 0.8 $ magnitudes. We observe no correlation between continuum flux and \ion{Mg}{2} line flux. This result resolves a controversy in the literature regarding the relationship of the continuum flux and \ion{Mg}{2} line flux in BP Tauri. We conclude that the bulk of the \ion{Mg}{2} in BP Tauri is not emitted below the accretion shock. Therefore, it may be emitted by the accretion column, the stellar wind and/or magnetospheric flaring. We do not observe any correlation between color temperature and the average value of the continuum. 

Within the context of the magnetospheric accretion paradigm, we have used models by CG98 to obtain the energy flux into the star due to the accretion process, and the spot size for each of the spectra of BP Tauri. We assume a constant inner truncation radius for the disk of $5\ R_*$. With 45 spectra, we find an average energy flux of $5\ 10^{11} {\rm{ergs\ cm^{-2}\ sec^{-1}}}$ and an average projected spot size of $4.4\ 10^{-3}$ times the area of one hemisphere of the star. The latter result coincides with previous observational determinations of spot sizes. From the energy flux and the spot size we obtain the accretion rate times $\cos \theta$, where $\theta$ is the average inclination angle of the region that produces the emission. The average of this projected accretion rate is $1.6\ 10^{-8} \msun \ \rm{yr^{-1}}\pm 30\%$, consistent with values found by other researchers. This values are weakly sensitive to the value of the underlying star used obtain the continuum excess. No correlation is observed between the flux in \ion{Mg}{2} line and the energy flux, the accretion rate or spot size, affirming the conclusion from the previous paragraph.

The nature of the data and the models is such that we cannot say anything about intrinsic variations in the accretion rate or the real (i.e., deprojected) spot size, but we conclude that the dispersion in their ratio is $\dot{M}/f\sim 50\%$, consistent with the errors. 

We observe a correlation between the ultraviolet flux and the accretion rate, similar to correlations found by other researchers between the U-band flux and the accretion rate: ${\log (L_{acc}/ \lsun)}={(1.1\pm0.1)*\log (L_{2900 \rm{\AA}}/ \lsun)-(0.5\pm0.05)}$. This correlation is implicit in the models by CG98. Also, we observe a correlation between the accretion rate ${\dot M}\cos\theta$ and the projected spot size $f$: ${\dot M}\cos\theta\  \alpha\ f^{0.5\pm0.05}$. Together, the two correlations imply that the UV flux in the surface of the star is proportional to the particle flux.

These results are significant, in spite of the large errors involved in the determinations of each quantity. These correlations stand in contrast to results obtained by Calvet \& Gullbring (1998). Using individual optical observations of 15 CTTSs, they concluded that the accretion rate increases linearly with spot size. A resolution of this discrepancy will have to wait until we have applied the magnetospheric accretion models to other stars.

Using a simulation, we conclude that the geometry of the system determined by the observer, the rotation axis and the inclination of the star can not explain the correlation between accretion rates and spot sizes. One must look at other more fundamental explanations.

Models of magnetospheric accretion by Ostriker \& Shu (1995) and Mahdavi \& Kenyon (1998) do predict that the spot size should increase with the accretion rate , but in both models, the predicted increase is much slower than what we found here ($f\ \alpha \ {\dot M}^2$). 

The correlation between accretion rate and spot size could be due to an unmodelled efficiency factor that determines how the spot is filled. Conversely, it indicates that the spot size is very sensitive to accretion rate. Future modeling should take into account the possibility of a non-dipole field for the star, oblique shocks, limb brightening and the effects of the geometry determined by the observer. The correlation between accretion rate and spot size puts a strong constraint in these models.

\acknowledgments

We would like to thank Nuria Calvet and Erik Gullbring for providing us with their models. Dr. Calvet in particular spent quite a bit of time helping us to understand the nature and strengths of their models. We are grateful to F. Shu and J. Bouvier for suggestions that led to the improvement of the text. Thanks are also due to Debi Howell-Ardila who edited the manuscript for language. We acknowledge the support of the NASA ADP program through grant \# NAG5-3471.


\clearpage
Figure 1: (a) Average of all the available spectra (solid line) and the normalized variance (dotted line). The feature near $3050\rm{\AA}$ is a reseaux mark. (b) Two sample spectra (lwp18976, solid line, lwp09256, dash-dot line) uncorrected by reddening. Notice the strong variation in slope and flux level. The position of some spectral features is indicated (Brown et al. 1984).

Figure 2: Mg II line flux vs. UV continuum flux. The continuum flux has been calculated from a box $50\rm{\AA}$ wide centered at $2900\rm{\AA}$. No correlation between the two quantities is observed, which means that the MgII line is not sensitive to the characteristics of the accretion spot.

Figure 3: Average of the logarithm of the flux versus the color temperature. Both have been calculated from the continuum fitting procedure (Section 4.2). No correlation is observed between the two quantities.

Figure 4: (a) Histogram for the calculated values of $f$. $\overline{f}=4.4\ 10^{-3}$ with a dispersion of $\sigma= 5\ 10^{-3}$. The errors are $\sim10\%$. The bar indicates the error at $f=0.0095$. (b) Histogram for the calculated values of $\log {\mathcal{F}}$, the logarithm of the energy flux into the star; $\overline{\log \mathcal{F}}=11.7$ and the dispersion is $\sigma=0.2$. The error is $\sim1$\%, as indicated by the bar centered in 11.4. With $\log {\mathcal{F}}$ and $f$ we calculate ${\dot M}\cos\theta$, the accretion rate uncorrected for projection effects. As shown in (c), $\overline{{\dot M}\cos\theta}=1.6\ 10^{-8} \msun \ \rm{yr^{-1}}$ and the dispersion is $\sigma \sim 0.8 \ 10^{-8}\msun/yr$. The errors are $\sim30$\%, as indicated by the bar for ${\dot M}\cos\theta=1.3\ 10^{-8} \msun \ \rm{yr^{-1}}$.

Figure 5: The projected accretion rate vs. the UV continuum flux. The UV continuum flux is taken from Table 1. This relation is similar to previously found relationships between the accretion luminosity and the U-band flux. 

Figure 6: (a) The logarithm of the accretion rate vs. $\log(f)$. The boxes indicate the extent of the errors. The correlation is such that ${\dot M}\cos(\theta)\ \alpha \ f^{0.5\pm0.05}$. (b) The correlation between  $\log {\mathcal{F}}$ and $\log(f)$ implied by (a) via Equation 1: ${\mathcal{F}}\ \alpha \ f^{-0.5}$. The solid line is what one would obtain from Equation 1, assuming a constant ${\dot M}\cos\theta=1.6\ 10^{-8} \msun \ \rm{yr^{-1}}$. (c) Same as (a) without the error boxes. The names indicate the positions of stars analyzed by CG98 using single optical observations (i.e. AA means AA Tauri, DK means DK Tauri and so on for all stars except GM Aur and UY Aur). In their data, some tendency can be seen to larger accretion spots with larger accretion rates.

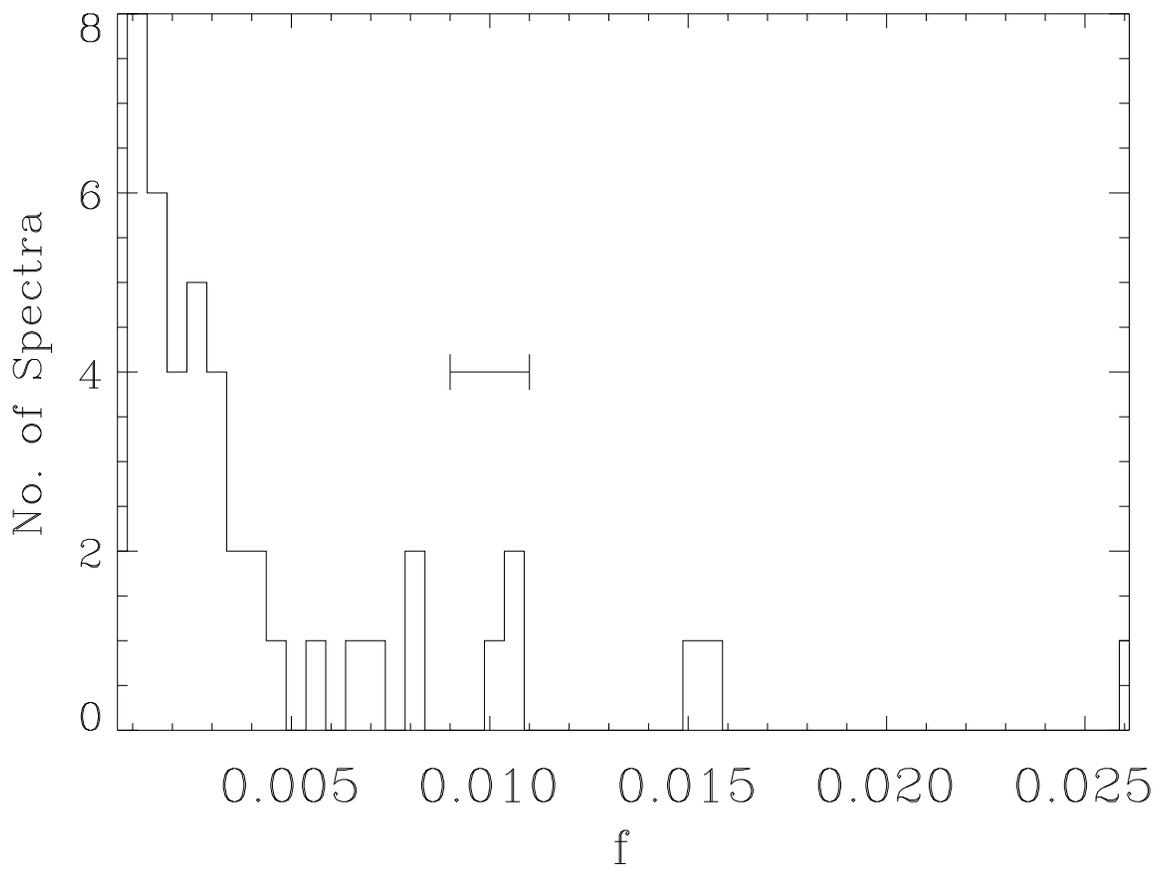
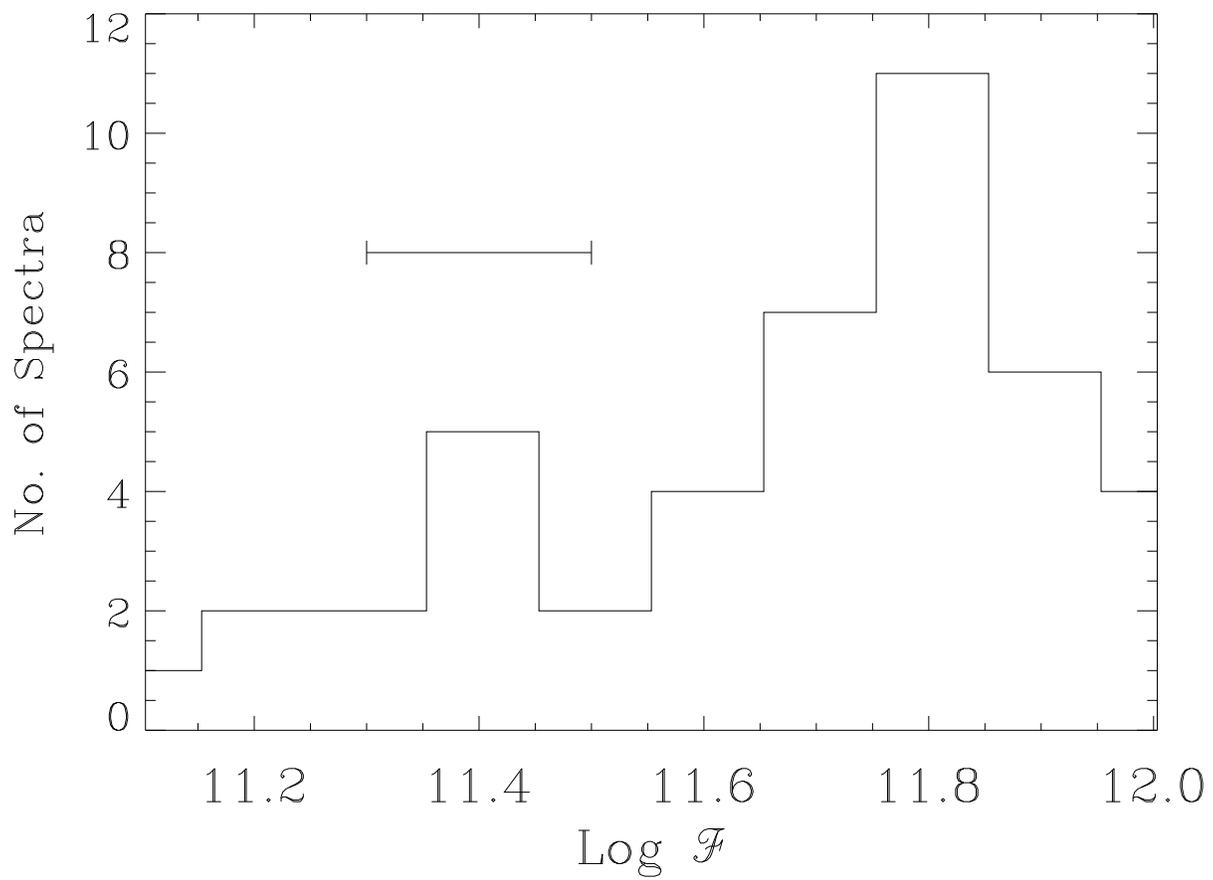
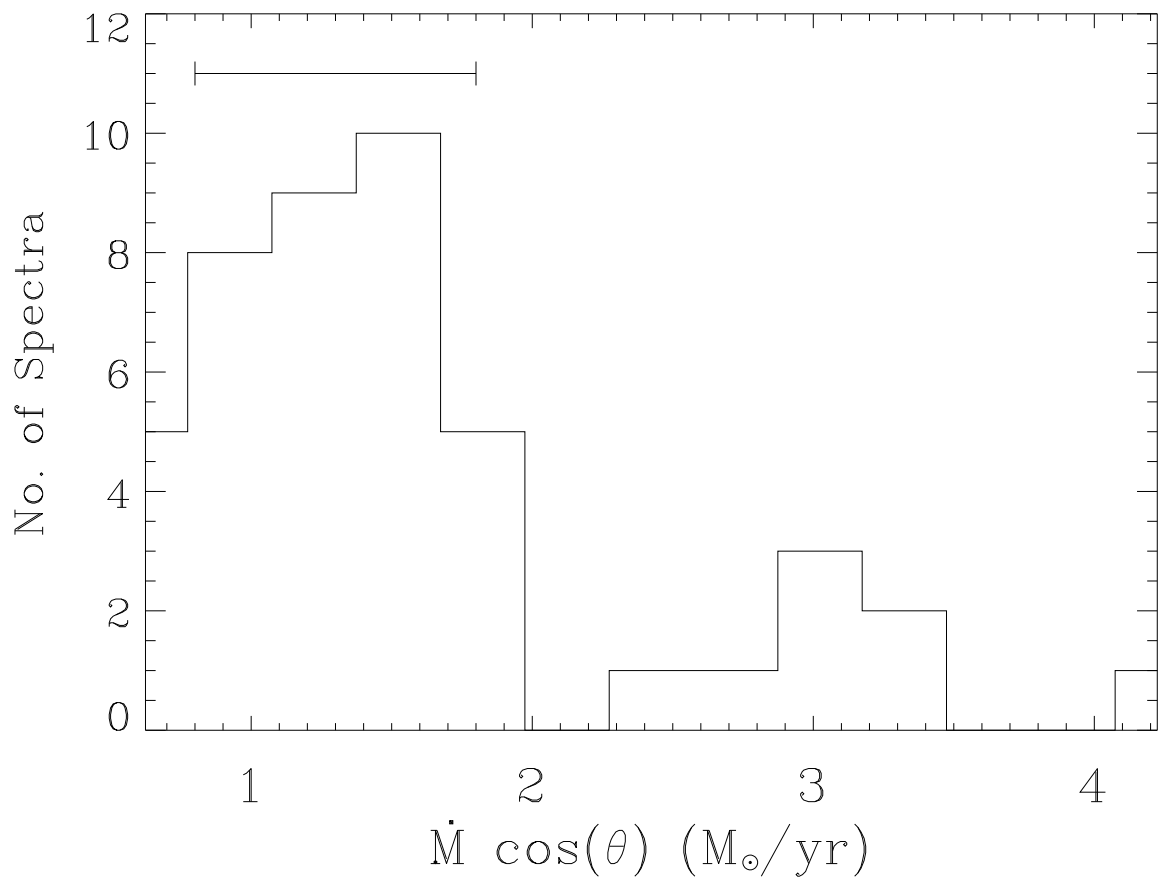
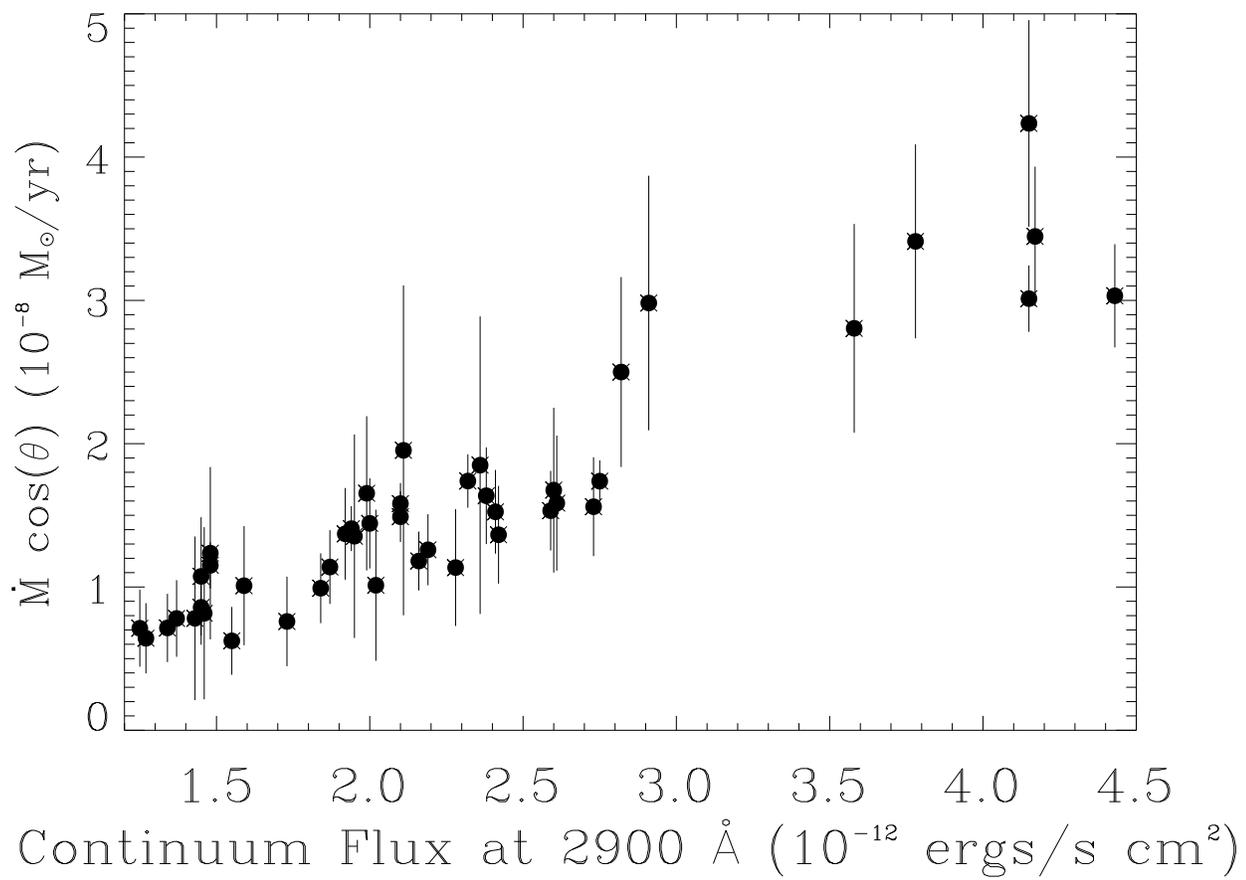
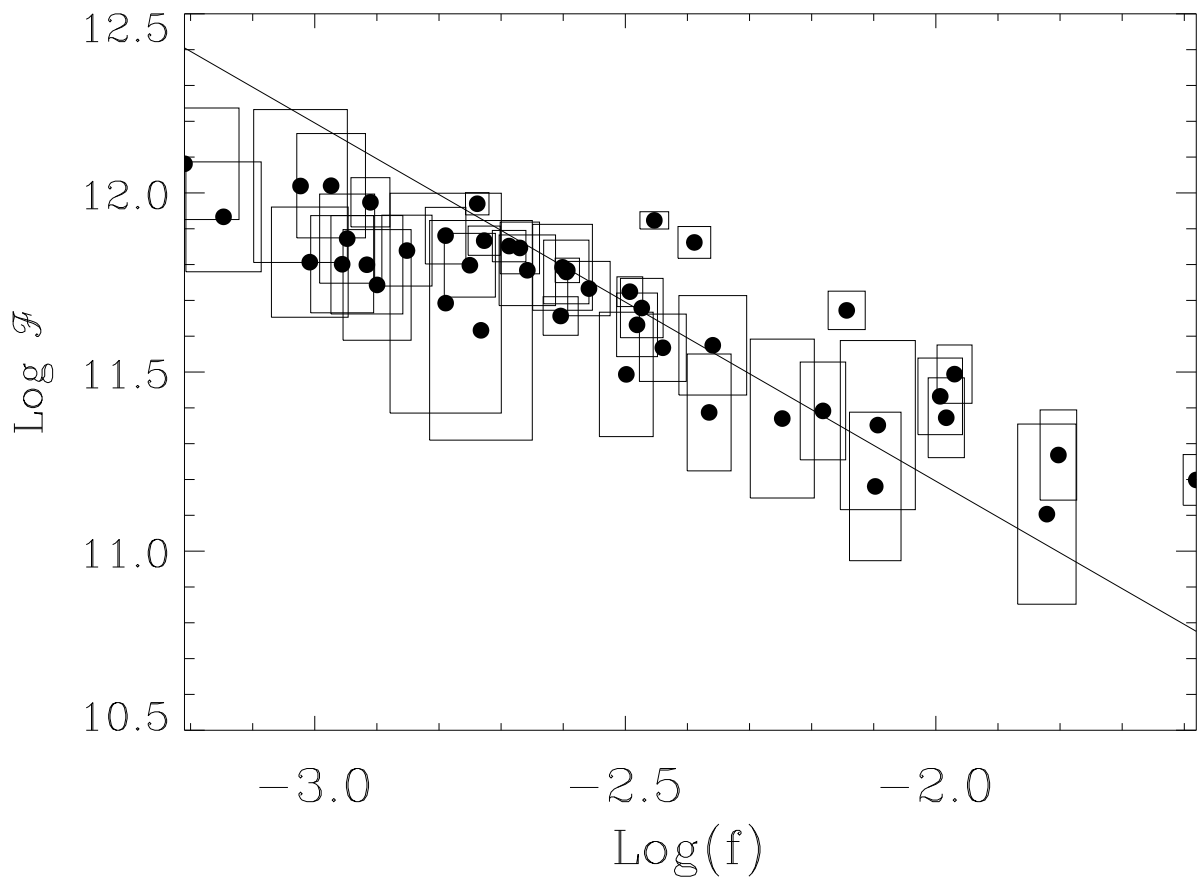
\end{document}


\pagestyle{empty}
\textheight=6in

\begin{deluxetable}{ccccccc}

\newcommand \msun{M_\odot}
\tablenum{3}
\tablecaption{Modeling the UV continuum}
\tablehead{
\colhead{Name} &
\colhead{T\tablenotemark{a}} & \colhead{Slope\tablenotemark{b}}& 
\colhead{Average\tablenotemark{c}}&
\colhead{$f$\tablenotemark{d}} &\colhead{$\log\mathcal{F}$\tablenotemark{e}} &
\colhead{$\rm{{\dot M}\cos\theta}$\tablenotemark{f}}}

\startdata

lwr04192 & $ 7300(400) $ & $2.74\ 10^{-4}$& $-13.502$ & $ 8.1(1.1)\ 10^{-3}$ & $ 11.4(0.2)$ & $ 1.8(1.0) $ \nl
lwr04192 & $ - $ & -& $ - $ & $ - $ & $ - $ \nl
lwr04249 & $ - $ & -& $ - $ & $ - $ & $ - $ \nl
lwr09309 & $ 8700(500) $ & $1.03\ 10^{-4}$& $-13.448$ & $ 2.5(0.3)\ 10^{-3} $ & $ 11.8(0.1) $ & $ 1.6(0.5) $ \nl
lwr11128 & - & -& - & - & -& - \nl
lwr11129 & - & -& $ -  $ & $- $ & $- $ \nl
lwr11131 & - & -& $ -  $ & $- $ & $-$ \nl
lwr14424 & - &-& $ -  $ & $- $ & $-$ \nl
lwr14425 & $ 7800(900) $ & $1.38\ 10^{-4}$& $-13.752$ & $ 1.6(0.3)\ 10^{-3} $ & $ 11.7(0.3) $ & $ 0.8(0.6)$ \nl
lwr15003 & - & -& $ - $ & $ - $ & $ - $ \nl
lwr15006 &$6900(300)$ & $3.18\ 10^{-4}$& $-13.738$ & $ 7.9(0.8)\ 10^{-3} $ & $ 11.2(0.2) $ & $ 1.2(0.6)$ \nl
lwr16945 & $7500(800)$& $1.66\ 10^{-5}$& $-13.787$ & $ 1.8(0.3)\ 10^{-3} $ & $ 11.6(0.3) $ & $ 0.8(0.6) $ \nl
lwr16962 &$9700(900)$ & $0.00         $& $-13.589$ & $ 9.4(2)\ 10^{-4}$ & $12.0(0.2)$ & $1.0(0.5)$ \nl
lwr16972 & - & -& $ - $ & $ - $ & $ - $ \nl
lwr16980 & - & - & $ - $ & $- $ & $- $ \nl
lwp09227 & $8000(500) $ & $9.09\ 10^{-5}$& $-13.835$ & $ 9.8(1)\ 10^{-4} $ & $ 11.8(0.2) $ & $ 0.6(0.2) $ \nl
lwp09228 & $ 8100(500)$ & $9.20\ 10^{-5}$& $-13.790$ & $ 1.1(0.1)\ 10^{-3} $ & $ 11.8(0.1) $ & $ 0.7(0.2)$ \nl
lwp09229 & $ 7900(200) $ & $1.87\ 10^{-4}$& $-13.638$ & $ 2.5(0.2)\ 10^{-3} $ & $ 11.6(0.1) $ & $ 1.1(0.2) $ \nl
lwp09230 & $ 9700(900) $ & $-2.58\ 10^{-5}$& $-13.707$ & $ 6.2(1)\ 10^{-4} $ & $ 12.1(0.2) $ & $ 0.8(0.3)$ \nl
lwp09231 & - & - & $ - $ & $ - $ & $ - $ \nl
lwp09254 & $ 8301(200) $ & $1.80\ 10^{-4}$& $-13.157$ & $ 7.2(0.5)\ 10^{-3} $ & $ 11.7(0.1) $ & $ 3.4(0.5)$\nl
lwp09255 & $ 9100(200) $ & $9.28\ 10^{-5}$& $-13.160$ & $ 4.1(0.2)\ 10^{-3} $ & $ 11.9(0.1) $ & $ 3.0(0.4)$ \nl
lwp09256 & $ 9230(100)$ & $1.18\ 10^{-4}$& $-13.170$ & $ 3.5(0.2)\ 10^{-3} $ & $ 11.9(0.1) $ & $ 3.0(0.2)$ \nl
lwp09257 & $ 7600(200) $ & $2.61\ 10^{-4}$& $-13.305$ & $ 1.0(0.08)\ 10^{-2} $ & $ 11.4(0.1) $ & $ 2.8(0.7) $ \nl
lwp09258 & $ 7800(200) $ & $2.41\ 10^{-4}$& $-13.200$ & $ 1.1(0.07)\ 10^{-2} $ & $ 11.5(0.1) $ & $ 3.4(0.7) $ \nl
lwp09283 & $ 8500(400)$ & $1.10\ 10^{-4}$& $-13.518$ & $ 2.2(0.2)\ 10^{-3} $ & $ 11.8(0.1) $ & $ 1.4(0.3)$ \nl
lwp09284 & $ 8900(200) $ & $9.89\ 10^{-5}$& $-13.472$ & $ 2.1(0.1)\ 10^{-3} $ & $ 11.8(0.1) $ & $ 1.5(0.2) $ \nl
lwp09285 & - & -& - & - & -   & - \nl
lwp09336 & $ 7400(400) $ & $2.16\ 10^{-4}$& $-13.708$ & $ 3.2(0.3)\ 10^{-3} $ & $ 11.5(0.2) $ & $ 1.0(0.4)$ \nl
lwp09337 & $ 7200(300) $ & $2.51\ 10^{-4}$& $-13.716$ & $ 4.3(0.3)\ 10^{-3} $ & $ 11.4(0.2) $ & $ 1.1(0.4) $ \nl
lwp09375 & $ 7400(200) $ & $2.66\ 10^{-4}$& $-13.365$ & $ 1.0(0.06)\ 10^{-2} $ & $ 11.4(0.1) $ & $ 2.5(0.7)$ \nl
lwp09376 & $ 7200(100) $ & $3.37\ 10^{-4}$& $-13.221$ & $ 2.6(0.1)\ 10^{-2} $ & $ 11.2(0.1) $ & $ 4.2(0.7) $ \nl
lwp09416 & - & -& $  - $ & $- $ & $- $            \nl
lwp09418 & $ -$ & -& - & - & - & $ - $ \nl
lwp14276 & $ 8900(200)$ & $8.14\ 10^{-5}$& $-13.490$ & $ 1.9(0.1)\ 10^{-3} $ & $ 11.7(0.1) $ & $ 1.4(0.2) $ \nl
lwp14791 & $ 8700(100) $ & $1.33\ 10^{-4}$& $-13.465$ & $ 2.5(0.1)\ 10^{-3} $ & $ 11.8(0.1) $ & $ 1.6(0.1) $ \nl
lwp16660 & - & -& $ -       $ & $   -   $ & $- $ \nl
lwp17185 & $ 9600(200) $ & $4.16\ 10^{-5}$& $-13.373$ & $ 1.8(0.1)\ 10^{-3} $ & $ 12.0(0.1) $ & $ 1.7(0.1)$ \nl
lwp18976 & $ 8400(200) $ & $1.58\ 10^{-4}$& $-13.437$ & $ 3.2(0.1)\ 10^{-3} $ & $ 11.7(0.1) $ & $ 1.7(0.2)$ \nl
lwp22194 & $ 7300(300) $ & $2.62\ 10^{-4}$& $-13.539$ & $ 6.5(0.5)\ 10^{-3} $ & $ 11.4(0.1) $ & $ 1.6(0.5)$ \nl
lwp22195 & $9000(400) $ & $6.24\ 10^{-5}$& $-13.528$ & $ 1.6(0.1)\ 10^{-3} $ & $ 11.8(0.1)$ & $ 1.3(0.2)$ \nl
lwp22211 & $ 8600(400) $ & $1.06\ 10^{-4}$& $-13.457$ & $ 2.5(0.2)\ 10^{-3} $ & $ 11.8(0.1) $ & $ 1.6(0.3) $ \nl
lwp22212 & $ 9500(800) $ & $-1.75\ 10^{-5}$& $-13.540$ & $ 1.1(0.1)\ 10^{-3} $ & $ 12.0(0.1) $ & $ 1.1(0.4)$ \nl
lwp22213 & $7800(400) $ & $1.88\ 10^{-4}$& $-13.469$ & $ 4.4(0.5)\ 10^{-3} $ & $ 11.6(0.1) $ & $ 1.7(0.6) $ \nl
lwp22226 & $8600(600) $ & $2.65\ 10^{-5}$& $-13.818$ & $ 7.1(1)\ 10^{-4} $ & $ 11.9(0.2) $ & $ 1.7(0.2)$ \nl
lwp22227 & $ 8000(300) $ & $1.79\ 10^{-4}$& $-13.530$ & $ 3.3(0.2)\ 10^{-3} $ & $ 11.6(0.1) $ & $ 0.6(0.3)$ \nl
lwp22228 & $ 7800(300) $ & $2.10\ 10^{-4}$& $-13.575$ & $ 3.6(0.3)\ 10^{-3} $ & $ 11.6(0.1)$ & $ 1.4(0.3)$ \nl
lwp22238 & $ 7300(400) $ & $2.63\ 10^{-4}$& $-13.625$ & $ 5.7(0.7)\ 10^{-3} $ & $ 11.4(0.2) $ & $ 1.4(0.7) $ \nl
lwp22239 & $ 9400(400) $ & $2.24\ 10^{-5}$& $-13.535$ & $ 1.2(0.1)\ 10^{-3} $ & $ 11.9(0.1) $ & $ 1.2(0.2) $ \nl
lwp22240 & $ 8600(400) $ & $ 1.03\ 10^{-4}$& $-13.591$ & $ 1.8(0.1)\ 10^{-3}$ & $ 11.8(0.1) $ & $ 1.1(0.3) $ \nl
lwp22243 & $ 8600(500) $ & $5.73\ 10^{-5}$& $-13.693$ & $ 1.1(0.1)\ 10^{-3} $ & $ 11.8(0.1) $ & $ 0.9(0.3) $ \nl
lwp22244 & - & -& -& -  & - & - \nl
lwp22245 & $ 8000(500) $ & $8.39\ 10^{-5}$& $-13.803$ & $ 1.2(0.1)\ 10^{-3} $ & $ 11.7(0.2) $ & $ 0.7(0.3)$ \nl
lwp22248 & $ 8300(500) $ & $1.17\ 10^{-4}$& $-13.755$ & $ 1.2(0.1)\ 10^{-3} $ & $ 11.8(0.2) $ & $ 0.8(0.3) $ \nl
lwp22249 & $ 7200(200) $ & $3.24\ 10^{-4}$& $-13.329$ & $ 1.6(0.1)\ 10^{-2} $ & $ 11.3(0.1) $ & $ 3.0(0.9)$ \nl
lwp22261 & $ 8200(300) $ & $3.00\ 10^{-4}$& $-13.467$ & $ 3.4(0.3)\ 10^{-3} $ & $ 11.7(0.1) $ & $ 1.6(0.3) $ \nl
lwp22262 & $8900(300) $ & $8.56\ 10^{-5}$& $-13.454$ & $ 2.1(0.1)\ 10^{-3} $ & $ 11.8(0.1) $ & $ 1.5(0.3)$ \nl
lwp22263 & $ 8400(300) $ & $1.38\ 10^{-4}$& $-13.486$ & $ 2.7(0.2)\ 10^{-3} $ & $ 11.7(0.1) $ & $ 1.5(1.2)$ \nl
lwp22276 & $ 7000(300)$ & $3.24\ 10^{-4}$& $-13.557$ & $ 1.5(0.01)\ 10^{-2} $ & $ 11.1(0.2) $ & $ 1.9(0.3) $ \nl
lwp22277 & - & -& -& - & - & - \nl
lwp22278 & $ 8700(400) $ & $8.28\ 10^{-5}$& $-13.642$ & $ 1.4(0.1)\ 10^{-3} $ & $ 11.8(0.1) $ & $ 1.0(0.2) $ \nl

 \enddata
\tablecomments{Empty positions in the table indicate spectra that are so noisy that a determination of the continuum over a sizable wavelength range is impossible. Numbers in parenthesis indicate errors}
\tablenotetext{a}{Color temperature in K}
\tablenotetext{a}{Slope of the logarithm of the fitted flux.}
\tablenotetext{b}{Average of the logarithm of the fitted flux.}
\tablenotetext{c}{Size of the accretion spot as fraction of the observed area of the star.}
\tablenotetext{d}{Logarithm of the energy flux}
\tablenotetext{e}{Projected accretion rate. Units: $10^{-8}\rm{\msun/yr}$}

\end{deluxetable}
\normalsize